\documentclass{svproc}

\usepackage{url}
\usepackage{graphicx}
\usepackage{amsmath}
\usepackage{hyperref}

\newcommand{\Lagr}{\mathcal{L}}
\def\Nbub{\langle {\cal N}\rangle}

\begin{document}
\mainmatter           
\title{Higgs vacuum metastability in $R+R^2$ gravity}
\titlerunning{Higgs vacuum metastability in $R+R^2$ gravity}
\author{Andreas Mantziris\inst{1, 2}}
\authorrunning{Andreas Mantziris} 
\tocauthor{Andreas Mantziris}
\institute{Department of Physics, Imperial College London, London, SW7 2AZ, United Kingdom
\and
Faculty of Physics, University of Warsaw, ul. Pasteura 5, 02-093 Warsaw, Poland \\
Email: \email{andreas.mantziris@fuw.edu.pl}}

\maketitle

\begin{abstract}
Experimental data suggest that the Higgs potential has a lower ground state at high field values. Consequently, decaying from the electroweak to the true vacuum nucleates bubbles that expand rapidly and can have dire consequences for our Universe. This overview of our last study \cite{Mantziris:2022fuu} regarding the cosmological implications of vacuum metastability during Starobinsky inflation was presented at the HEP2023 conference. Following the framework established in \cite{Mantziris:2020rzh}, we showcased our state-of-the-art lower bounds on the non-minimal coupling $\xi$, which resulted from a dedicated treatment of the effective Higgs potential in $R+R^2$ gravity. The effects of this consideration involved the generation of destabilising terms in the potential and the sensitive dependence of bubble nucleation on the last moments of inflation. In this regime, spacetime deviates increasingly from de Sitter and thus the validity of our approach reaches its limit, but at earlier times, we recover our result hinting against eternal inflation.
\end{abstract}

\section{Introduction}

\subsection{The metastability of the Higgs vacuum}

\begin{figure}[h]
\centering
\includegraphics[width=1\linewidth]{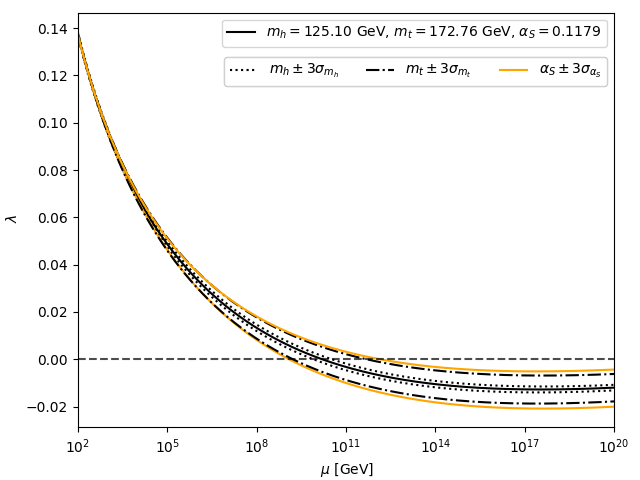}
\caption{The ``running'' of the Higgs self-interaction $\lambda$ with the renormalization scale $\mu$ according to the experimentally measured values of the Higgs mass $m_h$, top quark mass $m_t$, and the strong coupling $a_S$ to $3 \sigma$ variance \cite{ParticleDataGroup:2020ssz} .}
\label{fig:lambda}
\end{figure}

\begin{figure}[h]
\centering
\includegraphics[width=0.7\linewidth]{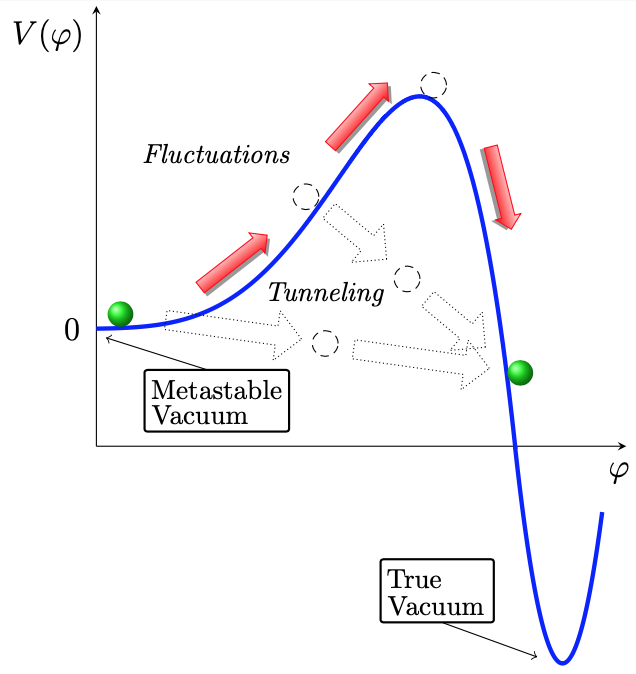}
\caption{The metastability of the vacuum state of a scalar field $\varphi$ in a potential $V(\varphi)$, where it can decay to the true vacuum by fluctuating over, tunnelling through, or a combination of both mechanisms \cite{Markkanen:2018pdo}.}
\label{fig:double_well}
\end{figure}

The data from collider experiments regarding the parameters of the Standard Model (SM) of particle physics \cite{ParticleDataGroup:2020ssz} indicate that the quartic coupling $\lambda$ of the Higgs field $h$ changes sign at high energies. The evolution of the self-coupling with the energy scale $\mu$ is formally computed by its $\beta$-function, where bosons and fermions contribute with opposing signs \cite{Rajantie:2018tqm}. The Higgs boson and the top quark, being the heaviest ones, dominate and seem to balance each other, resulting in the smooth function shown in Fig. \ref{fig:lambda}. The sign change of $\lambda$ leads to the development of a lower ground state than the electroweak (EW) vacuum in the Higgs potential
\begin{align}
  V_{\rm H}(h, \mu) = \frac{\lambda(\mu)}{4} h^4 \, + \dots \, .
  \label{eq:Vquartic}
\end{align} 
This realisation renders the EW vacuum metastable, since the Higgs field does not reside currently in the lowest energy state and is therefore prone to decay to its true vacuum \cite{Lancaster:2014pza,DeLuca:2022cus}. This process can take place quantum mechanically by tunnelling through the potential barrier, classically by fluctuating over it, or a combination of both, as shown in Fig. \ref{fig:double_well}.

The survival of the Higgs vacuum during our cosmological evolution can be utilised to impose significant constraints on fundamental physics in a manner that would enable this long-lasting metastability \cite{Markkanen:2018pdo}. In our studies \cite{Mantziris:2022fuu,Mantziris:2020rzh}, we adopted a minimal model of the early universe consisting of the SM and cosmological inflation in order to obtain vacuum decay bounds on the Higgs curvature coupling $\xi$. This coupling is the last missing renormalizable parameter of the SM, since it is impossible to probe it with accelerators in the flat spacetime of our late Universe, and thus can only be constrained by cosmological studies. The non-minimal coupling appears in the Higgs potential at tree-level
\begin{align}
  V_{\rm H}(h, \mu, R) = \frac{\xi(\mu)}{2}R h ^2 + \frac{\lambda(\mu)}{4} h^4 + \dots \, ,
  \label{eq:Vtree}
\end{align}
where $R$ is the Ricci scalar that quantifies the curvature of spacetime \cite{Markkanen:2017dlc}, to ensure the renormalizability of the theory on a curved spacetime \cite{Callan:1970ze}. Therefore, assuming that vacuum decay was sufficiently suppressed during inflation constrains $\xi$ accordingly.

\subsection{Vacuum decay and bubble nucleation}

After the Higgs vacuum decays to its true minimum at a particular point in spacetime, the surrounding field follows in a runaway process that leads to the formation of a true-vacuum bubble, that expands with a velocity approaching the speed of light \cite{Markkanen:2018pdo}. We cannot safely tell anything specific about the exotic physics of the bubble interior, but it is assumed, within the SM context, that the violent decay process results in the interior collapsing into a singularity \cite{Tetradis:2023fnu}. Given that the Higgs field is in the metastable EW state at the moment, no bubbles could have nucleated within our cosmological history. This argument enables us to constrain fundamental physics, by demanding that the potential should have been ``metastable enough'' in order to survive. Quantitatively, this means that the expectation value of the number of bubbles $\mathcal{N}$ has to obey 
 \begin{align}
        \Nbub \leq 1 \, ,
        \label{eq:constraint}
 \end{align}
in order to be compatible with observations \cite{Markkanen:2018pdo}. This quantity is given by the integral of the decay rate per spacetime volume $\Gamma$ over the past lightcone
 \begin{align}
     \Nbub = \int_{\substack{ \rm past }} d^4x \sqrt{-g} \Gamma(x) \, ,
    \label{eq:Nbubs}
    \end{align}
where $g$ is the determinant of the spacetime metric $g_{\mu \nu}$.

The early universe provides a fruitful context for these considerations, since various mechanisms could have enhanced the decay rate at such high scales \cite{Eichhorn:2018cyr,Cruz:2022ext,Vicentini:2022pra,Lebedev:2021xey}, and the apparent curvature of spacetime means the non-minimal term could have prominent effects on the stability of the EW vacuum. Therefore, we are motivated to study vacuum decay in the context of inflation, where spacetime was significantly curved. During inflation, the Universe undergoes a phase of exponential expansion due to the slow roll of a scalar field, i.e. the inflaton $\phi$, in an approximately flat potential \cite{Liddle:2000cg}. The duration of inflation is measured for convenience by $e$-foldings, defined as 
\begin{align}
N  \equiv \mathrm{ln} \left( \frac{a_{\rm inf}}{a}\right) \, ,
\end{align}
with $a$ corresponding to the scale factor and the index ``inf'' denoting the end of inflation. In order to comply with observations, a minimum number of approximately 60 $e$-foldings is required \cite{Liddle:2000cg}, meaning that
\begin{align}
N_{\rm start} \geq 60 \, .
\end{align}
Expressing Eq. (\ref{eq:Nbubs}) in the inflationary context results in 
\begin{align}
    \Nbub =  \frac{4\pi}{3}\int_0^{N_{\mathrm{start}}} dN \left( \frac{a_{\mathrm{inf}} \left(\eta_0-\eta\left(N\right)\right)}{e^{N}} \right)^3 \frac{\Gamma(N)}{H(N) } \, ,
    \label{eq:Nbub}
\end{align}
where we are integrating backwards in time from $N_{\rm inf} =0$ to $N_{\rm start}$, $\eta_0 - \eta(N)$ is the the comoving radius of the spacetime volume in terms of the conformal time $\eta$ between today and $N$ $e$-folds before the end of inflation, and 
\begin{align}
H  \equiv \frac{\dot{a}}{a} = \frac{da}{dt} \frac{1}{a}
\end{align} 
is the Hubble rate that measures the acceleration of the expansion.

Hence, we identify two independent calculations that coalesce into a comprehensive computation of lower bounds on the Higgs curvature coupling, 
\begin{align}
\xi \geq \xi_{\Nbub=1} \, ,
\end{align}
by imposing the requirement given by Eq. (\ref{eq:Nbubs}) on the integral in Eq. (\ref{eq:Nbub}). One stream involves the calculation of the relevant cosmological quantities with respect to the lightcone volume, i.e. the scale factor, conformal time, and Hubble rate. The dynamic evolution of these quantities is calculated beyond the simplifying slow-roll approximation and depends on the choice of the inflationary model, the definition of the end of inflation, and its total duration \cite{Mantziris:2020rzh}. In our case, we chose to consider \textit{Starobinsky Inflation} \cite{Starobinsky:1979ty}, due to its minimality and favourable compliance with observational data \cite{Planck:2018jri}, with the potential
\begin{align}
V_{\rm I}(\phi) = \frac{3 M^2 M_P^4}{4} \left(1-e^{-\sqrt{\frac{2}{3}}\frac{\phi}{M_P}} \right)^2 \, ,
\label{eq:Vstar}
\end{align} 
where $M=1.1 \times 10^{-5}$ is a dimensionless constant fixed by the CMB anisotropies and $M_P=2.435 \times 10^{18}$ GeV is the reduced Planck mass. At the same time, it is necessary to compute the time-dependent decay rate $\Gamma$ in de Sitter (dS) space according to the shape of the effective Higgs potential during inflation. We utilise the Hawking-Moss (HM) \cite{Hawking:1981fz} instanton solution for the estimation of $\Gamma$, 
\begin{align}
	\Gamma \approx \left(\frac{R}{12}\right)^2 e^{-\frac{384 \pi^2  \Delta V_{\rm H} }{R^2}} \, ,
\end{align} 
where $ \Delta V_{\rm H}$ is the height of the potential barrier between the two vacua, since this is the dominant process in the high Hubble scale regime during the inflationary epoch. For commentary on the comparison between the HM, the Coleman-de Lucia, and the stochastic formalism approaches, we refer the reader to \cite{Mantziris:2022ifn}.

\section{The effective Higgs potential in curved spacetime}

\subsection{Loop corrections and renomalisation group improvement} \label{sec:RGI}

Incorporating loop corrections to the tree-level potential (\ref{eq:Vtree}) leads to
\begin{align}
       V_{\rm H}(h, \mu, R) = \frac{\xi}{2} R h^2 + \frac{\lambda}{4} h^4 + \frac{\alpha}{144} R^2 + \Delta V_{\rm loops} \, ,
       \label{eq:Vloop}
 \end{align}
where the Higgs mass term has been neglected (because it is negligible compared to the high Hubble scales during inflation), a subdominant gravitational term is radiatively generated, and a loop term from the entire SM particle spectrum is included, consisting of 3-loop flat space corrections and 1-loop dS corrections
\begin{align}
     \Delta V_{\rm loops} = \frac{1}{64\pi^2} \sum\limits_{i=1}^{31}\bigg\{ n_i\mathcal{M}_i^4 \bigg[\log\left(\frac{|\mathcal{M}_i^2 |}{\mu^2}\right) - d_i \bigg] +\frac{n'_i R^2}{144}\log\left(\frac{|\mathcal{M}_i^2 |}{\mu^2}\right)\bigg\}  \, ,
     \label{eq:DeltaV}
\end{align}
where $\mathcal{M}_i (h, \mu, R)$ is the effective mass, and $n_i, \, d_i, \, n_i'$ are fixed parameters \cite{Markkanen:2018bfx}. 

We wish to simplify (\ref{eq:Vloop}) and eliminate its scale dependence since it is not physical, but only an arbitrary dimensionful constant \cite{Lancaster:2014pza}. Whether the perturbative expansion (\ref{eq:DeltaV}) converges safely is sensitive to the chosen value of $\mu$, but no singular option guarantees this for all field values \cite{Mantziris:2022ifn}. The usual approximation 
\begin{align}
\mu = h
\end{align}
is inadequate in this scenario, since it neglects the dominant contribution from curvature \cite{Markkanen:2018pdo}. To address this, the function
\begin{align}
\mu^2 = a h^2 + b R \, ,
\end{align}
was proposed in \cite{Herranen:2014cua}, with $a,b$ being fitted parameters. Employing this scale choice means that direct loop corrections do not entirely cancel out and must be incorporated into the effective potential \cite{Mantziris:2022ifn}. In our case, we utilise the most optimal and field-dependent scale choice via the procedure of Renormalization Group (RG) improvement, where $\mu=\mu_*(h,R)$ is the solution of the equation
\begin{align}
\Delta V_{\rm loops} (h, \mu_*, R) = 0 \, ,
\end{align}
resulting in the RG improved effective Higgs potential \cite{Mantziris:2020rzh}
\begin{align}
    V_{\rm H}^{\rm RGI}(h, R) = \frac{\xi(\mu_*(h, R))}{2}  R h ^2 +\frac{\lambda(\mu_*(h, R))}{4}  h^4 + \frac{\alpha(\mu_*(h,R))}{144} R^2 \, .
    \label{eq:Veff}
\end{align}

\subsection{Embedding the effective potential in $R+R^2$ gravity}

In the case of Starobinky inflation, which arises naturally by including a quadratic curvature term in the gravitational action \cite{Starobinsky:1979ty}, embedding the entire SM in $R+R^2$ gravity makes the computation of the RG improved effective potential more demanding. The action of the Higgs field in the Jordan frame, where gravity is modified by the addition of the geometric $R^2$-term to the Einstein-Hilbert action of General Relativity (GR), is given by
\begin{align}
    S = \int d^4x \sqrt{-g_J} \left[ \frac{M_P^2}{2} \left( 1 - \frac{\xi h^2}{M_P^2} \right) R_J + \frac{1}{12 M^2}R_J^2 +  \frac{1}{2} g_J^{\mu \nu} \partial_{\mu} h\partial_{\nu} h  - \frac{\lambda}{4}h^4 \right] \, ,
    \label{eq:action}
\end{align}
where $J$ denotes the value of each quantity in the Jordan frame. After a Weyl transformation to the Einstein frame 
\begin{align}
g_{J \, \mu \nu} = e^{- \sqrt{\frac{2}{3}}\frac{\phi}{M_P}} g_{\mu \nu} \, ,
\end{align} 
where gravity behaves according to GR, an additional scalar field $\phi$ in the matter sector carries the extra degree of freedom and plays the role of the inflaton. 

In order to be able to make appropriate comparisons and draw meaningful conclusions, we have to express the action (\ref{eq:action}) in a diagonal form. Firstly, we perform the field redefinition 
\begin{align}
\tilde{h} = e^{-\frac{1}{2} \sqrt{\frac{2}{3}}\frac{\phi}{M_P}} h \; , 
\end{align}
that enables the potential's RGI in the dS limit \cite{Mantziris:2022fuu} according to the prescription described in Section \ref{sec:RGI}. Two further redefinitions, one regarding the inflaton
\begin{align}
 \phi = \Tilde{\phi} - M_P \sqrt{\frac{3}{2}} \mathrm{ln} \left[ 1 - \left(\xi - \frac{1}{6}\right) \left( \frac{ \Tilde{h} }{M_P} \right)^2 \right] \, ,
    \label{eq:field-redef2}
\end{align}
and the other for the Higgs field 
\begin{align}
        \Tilde{h} \approx \rho \left[ 1 + \left(\xi - \frac{1}{6}\right)^2 \left( \frac{ \rho }{M_P} \right)^2 + \mathcal{O}(\rho^4) \right] \, .
        \label{eq:tildeh-to-rho}
\end{align}
result in the approximately canonical Lagrangian given by
\begin{align}
	\Lagr \approx  \frac{M_P^2}{2}R + \frac{1}{2}\partial_{\mu} \Tilde{\phi} \partial^{\mu} \Tilde{\phi} + \frac{1}{2} \partial_{\mu} \rho \partial^{\mu}  \rho   - \Tilde{U}(\Tilde{\phi}, \rho) \, .
\end{align}
The total potential $\Tilde{U}$ includes the Starobinsky potential (\ref{eq:Vstar})
\begin{align}
\Tilde{U}(\Tilde{\phi}, \rho) = V_{\rm I}(\Tilde{\phi}) +  m_{\rm eff}^2 (\Tilde{\phi}, \mu_*) \frac{\rho^2}{2} + \lambda_{\rm eff} (\Tilde{\phi}, \mu_*) \frac{\rho^4}{4} + \frac{\alpha (\mu_*)}{144} R^2 (\Tilde{\phi}) + \mathcal{O}\left(\frac{\rho^6}{M_P^2}\right) \, ,
\end{align}
and the other terms resemble the effective Higgs potential (\ref{eq:Veff}), with the addition of a Planck suppressed sextic term and the $\Tilde{\phi}$-dependent contributions to the effective quadratic and quartic couplings given by
\begin{align}
    m_{\rm eff}^2 &= \xi R + 3 M^2 M_P^2  \left(\xi - \frac{1}{6}\right) \left(1-e^{-\sqrt{\frac{2}{3}}\frac{\Tilde{\phi}}{M_P}}\right) e^{-\sqrt{\frac{2}{3}}\frac{\Tilde{\phi}}{M_P}} + \frac{ \left(\xi - \frac{1}{6}\right)}{M_P^2} \partial_{\mu} \Tilde{\phi} \partial^{\mu} \Tilde{\phi} \, , 
    \label{eq:meff} \\
    \lambda_{\rm eff} &= \lambda + \frac{3 M^2  \left(\xi - \frac{1}{6}\right)^2}{e^{2\sqrt{\frac{2}{3}}\frac{\Tilde{\phi}}{M_P}}} + \frac{4 \left[ \xi R + \Delta m^2_1 \right]  \left(\xi - \frac{1}{6}\right)^2}{M_P^2} +  \frac{4  \left(\xi - \frac{1}{6}\right)^3}{M_P^4} \partial_{\mu} \Tilde{\phi} \partial^{\mu} \Tilde{\phi} \,,
\end{align}
respectively, where $\Delta m_1^2$ corresponds to the middle term of $m^2_{\rm eff}$ in Eq. (\ref{eq:meff}). 

\begin{figure}[h!]
    \centering
\includegraphics[width=1\linewidth]{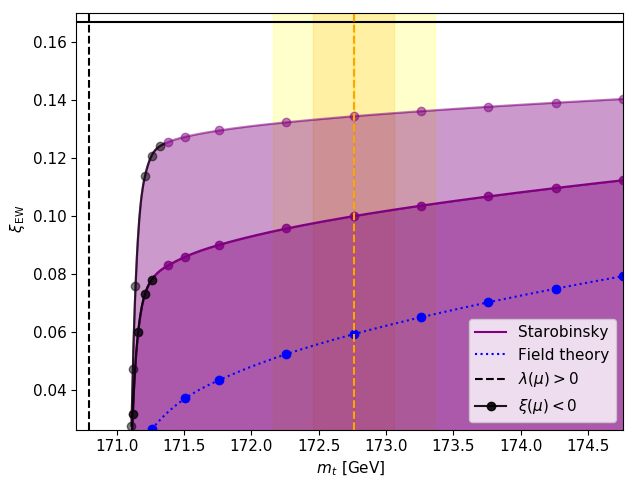}
\caption{Lower constraints on the Higgs curvature coupling at the electroweak scale $\xi_{\rm EW}$ for varying top quark mass $m_t$ (vertical dashed orange line: central value, shades: $\pm \sigma_{m_t}$, $\pm 2\sigma_{m_t}$). The parameter space below each curve is excluded, with the end of inflation set at $\dot{H}/H^2 = -1/4$ and $\dot{H}/H^2 = -1$ for the dark and light purple curves, respectively. The black parts highlight the value below which the $\xi$ turns negative as it runs. The field theory bounds \cite{Mantziris:2020rzh} shown by the dotted blue curve are included for comparison. The vertical dashed black line is at the threshold of $m_t$ below which $\lambda$ remains positive, and the horizontal black line shows the conformal point $\xi = 1/6$.}
\label{fig:bounds}
\end{figure}

The HM formalism is valid for $0<\xi_{\rm EW}<1/6$, meaning that the inflaton-dependent terms in Eq. (\ref{eq:meff}) are negative and thus act against the stabilising $\xi R$ term. Therefore, the lower constraints on $\xi$ are stronger in this case when compared to the field theory case of \cite{Mantziris:2020rzh}, since the Higgs curvature term needs to counter these additional terms in order to ensure the survivability of the metastable EW vacuum. The $\xi$-bounds derived with this prescription for varying top quark mass and with different definitions for the end of inflation are shown in Fig. \ref{fig:bounds}. In the $R+R^2$ context, we have obtained $\xi_{\rm EW} \geq 0.13$ in contrast to the field theory bound $\xi_{\rm EW} \geq 0.06$ for input values of $m_t = 172.76$ GeV, $m_h = 125.10$ GeV, and $\ddot{a}=0$ defining the endpoint of inflation.

In general, bubbles nucleate towards the end of inflation, but the additional terms in Eq. (\ref{eq:meff}) postpone bubble production. Spacetime departs from dS in this regime; however the decay rate $\Gamma$ and the effective RGI potential were calculated in the dS limit. Hence, if we adopt a more conservative assumption for the inflationary finale, we can quote $\xi$-bounds that are weaker but more reliable. For example, if we perform the same calculation assuming inflation lasts until $\frac{\ddot{a}}{a} = \frac{H^2}{2}$, the constraint on the non-minimal coupling is $\xi_{\rm EW} \geq 0.1$.

\section{Conclusions}

With this overview of our most recent investigation \cite{Mantziris:2022fuu}, we showcased a fruitful method for constraining fundamental physics via the synergy between SM physics and early universe cosmology. More specifically, we studied the experimentally suggested feature of Higgs vacuum metastability in the context of the observationally supported model of Starobinsky inflation. The novelty of our approach lies in the fact that we went beyond the dS description of the inflationary spacetime when calculating the relevant cosmological quantities, and we took into account the time dependence of the Hubble rate instead of setting it as a constant free parameter. In addition, we extended the range of similar studies \cite{Fumagalli:2019ohr,Li:2022ugn}, by improving also the calculation of the effective potential to 3-loops, with curvature corrections in dS to 1-loop, and a robust RGI prescription.

By employing the analytical and numerical techniques outlined in \cite{Mantziris:2022fuu}, we derived more robust $\xi$-bounds than the dS studies \cite{Mantziris:2020rzh,Markkanen:2018pdo}
        \begin{align}
            \xi_{\rm EW} \geq 0.1 > 0.06  \, ,
        \end{align}
through the computation of the most accurate RG improved effective Higgs potential at the time, which consists of additional destabilising terms. The evaluation of $\Gamma_{\rm HM}$ and the effective potential in the dS regime begin to break down when reaching the time of prominent bubble formation towards the end of inflation. Hence, it becomes increasingly crucial to start accounting for the reheating dynamics \cite{Kost:2021rbi,Figueroa:2021iwm}, if we are to achieve higher accuracy in the estimation of the Higgs curvature coupling. On the other hand, the effective potential matches the field theory scenario at earlier times, where the slow-roll approximation is valid, resulting in similar qualitative insights and indications against eternal inflation, akin to those found in \cite{Mantziris:2020rzh}.

\section*{Acknowledgements}

The author acknowledges the supervision of A. Rajantie and T. Markkanen, along with the support of an STFC PhD studentship, for the study \cite{Mantziris:2022fuu} that was presented in HEP2023 - $40^{\rm th}$ Conference on Recent Developments in High Energy Physics and Cosmology at the University of Ioannina.

\end{document}